\documentclass[a4paper,aps,prb,twocolumn,floatfix,showpacs]{revtex4}
\usepackage{multirow}
\usepackage{color}
\usepackage{hyperref}
\usepackage{url}
\usepackage[dvips]{graphicx}
\usepackage{amsmath}
\usepackage{graphicx,amscd,amsmath,amssymb,verbatim,subfigure}

\begin{document}

\title{Superconducting transition in Pb/Co nanocomposites: effect of Co volume fraction and external magnetic field}
\author{Y. T. Xing}
\author{H. Micklitz}
\author{W. A. Rodrigez}
\author{E. Baggio-Saitovitch}

\affiliation {Centro Brasileiro de Pesquisas F\'{i}sicas, Rua Dr.
Xavier Sigaud 150, Rio de Janeiro, 22290-180, RJ, Brazil}

\author{T. G. Rappoport} \affiliation{Instituto de
 F\'{\i}sica, Universidade Federal do Rio de Janeiro, Cx. P. 68528, 21941-972 , Rio de Janeiro, Brazil}

\date{\today}

\begin{abstract}

Pb films embedded with homogeneously distributed cobalt (Co) nanoparticles (mean size 4.5 nm) have been prepared. Previous transport investigations have shown that Co particles induce spontaneous vortices below the superconducting transition temperature (T$_{c}$) in zero external magnetic field. In this paper we study in detail the influence of the Co volume franction and an external magnetic field on the superconducting transition in such composites. The large difference in T$_c$-reduction between the as-prepared and annealed samples can be attributed to the different superconducting coherence lengths and the resulting different diameters of the spontaneous vortices in these samples.

\end{abstract}

 \maketitle


\section{introduction}

Composite materials with a nanometer scale structure, so-called nanocomposites, are an actual research field in solid state physics. In this way a hybrid of two different materials with conflicting properties can be artificially fabricated, which usually does not exist in nature. An example for such nanocomposites are systems made from superconducting (SC) and ferromagnetic (FM) materials. In such systems the interplay between the mutual exclusive cooperative phenomena SC and FM can be studied and many papers have been published on this topic\cite{AIBuzdin05, MLange03,JIMartin97, LNBulaevskii00, LRTagirov99, JEVillegas06,RLAiho03}. It was found that due to the interaction between SC and FM, the external magnetic field can increase the critical parameters of the superconductor. Other interesting phenomena, such as domain wall superconductivity\cite{ZRYang04}, hysteresis pinning\cite{APalau07}, etc, were also found in such kind of nanocomposites. There are mainly three types of SC/FM nanocomposites: SC and FM multilayers\cite{CMonton07}, FM decorations on tope of SC films and SC/FM granular systems. The first system is usually used to study the proximity effect in the SC/FM layers. The second one is usually used to study the formation of vortices and antivortices induced by the FM decorations, which can have the effect of critical parameter enhancement~\cite{MLange05,MJVanBael99}. In the last ones, namely, the granular hybrid systems the collective interactions between SC and FM has been studied. The above mentioned hysteresis pinning, for example, has been discovered in such a system~\cite{APalau07}. Very recently we reported about a SC/FM hybrid system containing single domain Co particles inside a SC Pb matrix~\cite{YTXing08}. The magnetic moments of the FM Co particles induce spontaneous vortices inside the SC matrix. The random network of such a spontaneous vortex phase has been studied without and with applied external magnetic field. In this paper we will report on the effect of external magnetic field on the superconducting transition temperature of such Pb/Co nanocomposites.

\section{Sample preparation}

The Pb/Co films were prepared by co-deposition of well-defined Co clusters and Pb atoms onto the sapphire substrate mounted onto the coldfinger of a rotatable, variable-temperature $^4$He cryostat. Sapphire was used as a substrate in order to have a good thermal contact between the sample and the coldfinger and having also an insulator which will not give any contribution to the transport measurements made on the film. Ag contacts for transport measurements have been deposited on the substrate by sputtering and were connected to the measurement system before closing the main chamber. The Co clusters were made in-beam by the so-called inert-gas-aggregation method with an Ar pressure of about 10$^{-1}$ mbar. A He-cryopump sitting between the cluster source and the main chamber absorbed the Ar gas. For that reason, only well-defined Co clusters and essentially no Ar atoms will enter the main chamber. The Pb atoms were thermally evaporated and deposited together with the Co clusters. Since it is well-known that island formation will prevent good quality Pb films if deposited at room temperature, the substrate was cooled down to $\sim$40 K during deposition which is low enough to get homogeneous Pb films, but, on the other hand, is high enough to prevent deposition of Ar atoms [p(Ar) at 40 K $\sim$ 10$^{-3}$ mbar] which have not been absorbed by the He-cryopump and, therefore, enter the main chamber together with the Co clusters. The angle between the matrix and the cluster beam is 45$^{o}$. Due to the different beam directions samples with different Co volume fraction could be made within one deposition process. The deposition rates were controlled by three quartz balances in order to monitor the rates at different positions of the substrate. The sample for microstructure study was deposited onto a carbon foil mounted in a Transmission Electron Microscope (TEM)-catcher.

After transport measurements on the as-prepared samples they were annealed at 300 K for one hour in order to decrease the lattice defects, formed in the samples due to the low-temperature deposition. The transport measurements have been repeated on the annealed samples in order to see the influence of lattice defects. The typical dimensions of the sample for transport and magnetic measurements were 10 mm $\times$ 3 mm $\times$ 100 nm. Transport properties in both zero and non-zero magnetic field were investigated in-situ with a split-coil superconducting magnet ( B $\leq$ 1.2 T). The detailed description of the experimental set-up and the advantage of the method can be found in the literature~\cite{SRubin98}. The magnetic measurements were performed with a Quantum Design MPMS-XL SQUID. The sample has been taken from the main chamber of the cluster source after finishing the transport measurements and immediately installed into the SQUID system in order to avoid oxidation. The external magnetic field is parallel to the surface of the sample in both the in-situ magnetotransport and ex-situ magnetic measurements. In this work we studied three samples with different Co cluster volume fraction: sample 1 with 2.7 vol\%, sample 2 with 3.1 vol\% and sample 3 with 3.7 vol\%.

\section {Results and discussion}

\begin{figure}[!ht]
    \centerline{\scalebox{.2}{\includegraphics{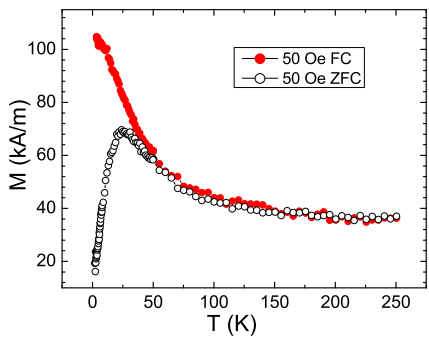}}}
    \caption{Magnetization as a function of temperature for the sample with about $3.7 vol\%$ Co. A 50 Oe external magnetic field was applied for both zero-field cooling and field cooling measurements. }\label{fig:MTcurves}
\end{figure}

The size of the Co nanoparticles was studied with a TEM and the picture can be found elsewhere\cite{YTXing08}. The Co nanoparticles are very
homogeneous in size and shape with a mean diameter of $\sim$ 4.5 nm. The superparamagnetic blocking temperature T$_b$ of the Co clusters has been determined in the usual way, namely, by measuring the magnetization of the sample in a warming-up process after cooling the sample down in either zero-field (zero-field cooling, ZFC) or in a magnetic field (field cooling, FC). A mean value of T$_b$ $\sim$ 25 K (corresponding to the peak temperature in the ZFC curves) is obtained from the experimental data shown in Fig. \ref{fig:MTcurves}. Measurement of the magnetization as a function of external magnetic field at 8 K, i. e. far below T$_b$, reveal the Co clusters to be in a ferromagnetic state\cite{YTXing08}.

\begin{figure}[!ht]
    \centerline{\scalebox{.2}{\includegraphics{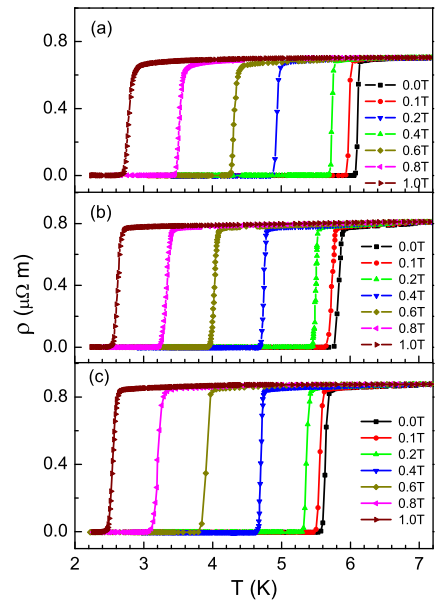}}}
    \caption{Resistivity as a function of temperature in different external magnetic fields for the as-prepared samples with (a) $2.7 vol \%$ Co (b) $3.1 vol \%$ Co and (c) $3.7 vol \%$ Co.}\label{fig:RT-AP}
\end{figure}

The resistivity measurements of all three as-prepared samples in zero field and for different external magnetic fields are shown in Fig. 2. One can see a quite sharp superconducting transition at T$_c$ with a slight decrease of T$_c$ with increasing Co volume fraction v and a large shift of T$_c$ caused by the external magnetic field. Here we do not show the as-prepared pure Pb sample because the difference of T$_c$ between the as-prepared and annealed pure Pb film (see below) is very small (about 0.1 K).

\begin{figure}[!ht]
    \centerline{\scalebox{.18}{\includegraphics{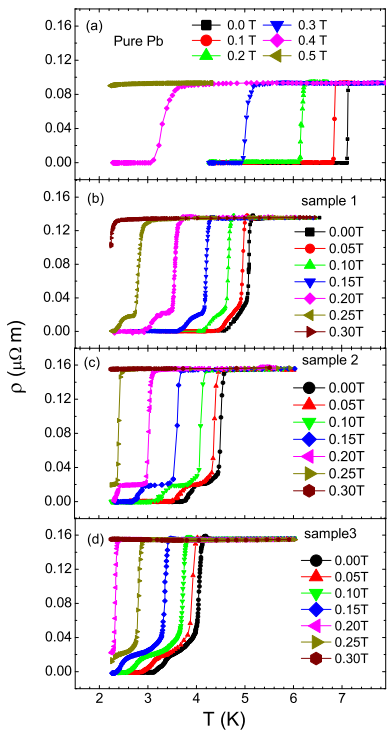}}}
    \caption{Resistivity as a function of temperature for different samples annealed at 300 K in different external magnetic fields. (a) pure lead, (b) with $2.7 vol\%$ Co, (c) with $3.1 vol\%$ Co and (d) with $3.7 vol\%$ Co.}\label{fig:RTcurves}
\end{figure}

Annealing the samples at 300 K has a drastic effect on the normal state resistivity ($\rho_N$) and the superconducting transition[see Fig. 3 (a)-(d)].  $\rho_N$ drops by a factor of $\sim$ 5. The reason for the large change in $\rho_N$ is the following: as-prepared samples, deposited at $\sim$ 40 K, will contain a large number of lattice defects, i.e. the Pb matrix will be in a highly disordered state. The electron mean free path, estimated from $\rho_N$ within the free electron model, is $\sim$ 0.6 nm, indicating that the Pb matrix may be even close to an amorphous state. Annealing at 300 K strongly reduces the number of lattice defects and increases the mean free path $l$ by a factor of $\sim$ 5, i.e. to $l \sim 3$ nm. The superconducting transition is shifted towards lower temperatures and the transition now occurs in two steps, i.e. the sharp transition seen in the as-prepared samples now has an ohmic region starting $\sim$ 1 K below the main transition occurring at T$_c$ and finally drops to zero resistivity $\sim$ 0.5 K lower. It should be mentioned that such a two step transition is not seen in the annealed pure Pb film [see Fig. 3 (a)]. The origin of the second transition has been discussed in detail in one of our previous papers~\cite{YTXing08}. It has been interpreted as a second-order phase transition in the spontaneous vortex phase from a vortex solid to a vortex liquid state. We will not focus on this point in the present paper but rather concentrate on the superconducting transition into the normal state occurring at T$_c$.

\begin{figure}[!ht]
    \centerline{\scalebox{.2}{\includegraphics{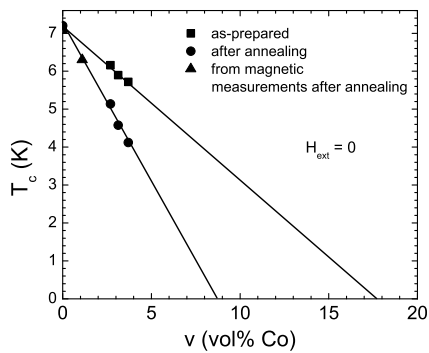}}}
    \caption{Critical temperatures, T$_c$, as a function of Co volume fraction with zero external magnetic field.}\label{fig:tc-co}
\end{figure}

We have plotted in Fig. 4 the dependence of T$_c$ from the Co volume fraction v for the three samples and the reference point (pure Pb). An additional T$_c$-point resulting from magnetic measurements on a sample containing $\sim$ 1 vol\% Co clusters has been added (see triangle in Fig. 4). The T$_c$-data are consistent with a linear dependence of T$_c$ with v. The straight lines drawn in Fig. 4 have a slope of dT$_c$/dv = - 0.43 K/vol\% Co for the as-prepared sample and dT$_c$/dv = - 0.85 K/vol\% Co for the sample annealed at 300 K.  The critical concentration for the disappearance of superconductivity is $\sim$17 vol\% for the as-prepared samples and $\sim$ 8 vol\% for the annealed samples. Both values are far below the value one would expect for non-magnetic particles. In Pb/Cu granular systems, for example, a critical volume fraction of $\sim$ 70 vol\% Cu has been found~\cite{ISternfeld05}. The T$_c$-reduction for granular SC systems with non-magnetic particles is due the proximity effect only, in the case of magnetic particles, however, an additional much stronger reduction occurs due to the formation of the above mentioned spontaneous vortices. The radius of these vortices is given by the superconducting coherence length $\xi$. For disordered superconductors, having a small electron mean free path $l$, the coherence length is given within the dirty limit to be $\xi \propto l^{1/2}$ or $\propto {\rho_N}^{-1/2}$. The change in $\rho_N$ going from the as-prepared to the annealed samples is 5.3 $\pm$ 0.5 for all three samples. For that reason $\xi$ should increase by a factor of 2.3 $\pm$ 0.1. The observed change of dT$_c$/dv going from the as-prepared to the annealed samples is $\sim$ 2, i.e. our  finding is that dT$_c$/dv scales with the coherence length or with the diameter of the spontaneous vortices.

\begin{figure}[!ht]
    \centerline{\scalebox{.17}{\includegraphics{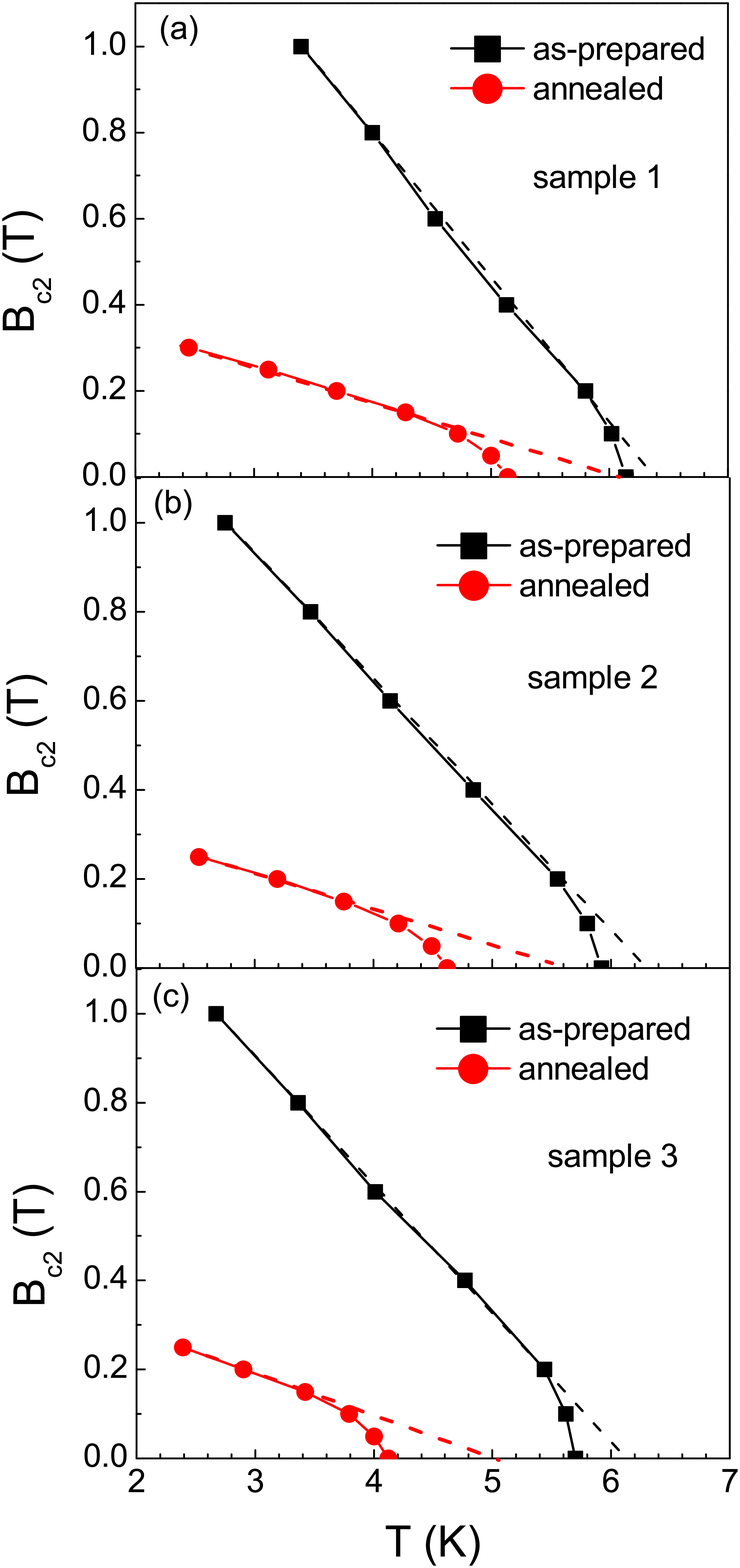}}}
    \caption {The on-set T$_c$ as a function of external magnetic field for the as-prepared and annealed samples with different Co $vol \%$.}\label{fig:TcH}
\end{figure}

Next we will discuss the T$_c$-reduction due to the applied external magnetic field. We have plotted in Fig. 5 the critical magnetic field B$_{c2}$ as a function of T for all three samples in the as-prepared as well as in the annealed state as obtained from Fig. 2 and 3. All B(T) curves show a curvature near T$_c$(B=0) which is not expected for a dirty superconductor. According to the work by Werthamer, Helfand and Hohenberg (so-called WHH-theory)~\cite{NRWerthamer66} there should be a linear dependence of B$_{c2}$(T) with T near T$_c$ with a slope given by
\begin{equation}
(dB_{c2}/dT)_{T = T_c} = (-4ek_B/\pi) N(E_F) \rho_N
\end{equation}
where N(E$_F$) is the density of states near the Fermi energy. We have drawn straight lines through the data points in Fig. 5 neglecting those for small external fields for obtaining (dB$_{c2}$/dT)-values for all samples in the as-prepared as well as in the annealed state. The ratio [(dB$_{c2}$/dT)$_{as-prepared}$ / (dB$_{c2}$/dT)$_{annealed}$] for all three samples is 4.2 $\pm$ 0.4 . If we compare this with the ratio ($\rho_N$)$_{as-prepared}$/($\rho_N$)$_{annealed}$ = 5.3 $\pm$ 0.5 (see above), we see that the change of (dB$_{c2}$/dT) due to annealing is somewhat smaller than that in $\rho_N$, but essentially in agreement with WHH-theory. Now we have to discuss why for small external magnetic fields the measured T$_c$-values are smaller than expected from the extrapolated straight lines in Fig. 5. Without external magnetic field the spontaneous vortices will form some kind of random network due to the random orientation of the Co particle magnetic moments. Applying an external magnetic field  above T$_c$ will align the magnetic moments [see hysteresis in Fig. 1 (b)] and, therefore, will also align the spontaneous vortices. For external fields of the order of 0.2 T the alignment of the magnetic moments  will be almost saturated and with external fields above this value the deviations of the measured data points from the extrapolated straight lines essentially disappear. The interpretation of our T$_c$-data for small external magnetic fields, therefore, is as follows: at small external magnetic fields the spontaneous vortex state develops from a random network to an aligned vortex state which somehow compensates the decrease of T$_c$ due to the external magnetic field; when all magnetic moments of the Co particles are aligned, the spontaneous vortex state is aligned and when the external magnetic field is further increased,  B$_c$(T) will follow the WHH-curve, i.e. will show a linear increase of B$_c$(T) with decreasing T.  The difference in T$_c$ between the measured T$_c$(B=0) values, i.e. the values for a random network of spontaneous vortices, and the extrapolated values for aligned spontaneous vortices is much larger for the annealed samples than for the as-prepared samples (see Fig.5), i.e. this difference is increasing with increasing coherence length or vortex diameter.\\

\section{Conclusion}

We have shown that the effect of both Co volume fraction as well of external magnetic field strongly depends if the samples are studied in the as-prepared (at low temperature) or in the annealed state. The reason for this is that the annealing strongly changes the electron mean free path and, as a consequence, the superconducting coherence length, which determines the diameter of the spontaneous vortices created by the magnetic moments of the Co particles. The reduction of T$_c$ due to Co particles is  much larger than that observed for non-magnetic particles in superconducting films, indicating that the influence of these spontaneous vortices on T$_c$ is much larger than the proximity effect. The T$_c$-reduction due to these spontaneous vortices seems to scale with the superconducting coherence length or diameter of these vortices. It would be interesting to see if theoretical calculations regarding the random network of spontaneous vortices could confirm our experimental result.


This work was partially supported by CAPES/DAAD cooperation program
and the Brazilian agencies CNPq, FAPERJ (Cientistas do Nosso Estado
and  PRONEX). H. Micklitz acknowledges FAPERJ
for financial support.



%
%
%
%
%
%
%
%
%
%

\end{document}